\documentclass{article}
\usepackage{spconf,amsmath,graphicx}
\usepackage{color}
\usepackage{hyperref}
\usepackage{subcaption}

\title{Optimized Power Normalized Cepstral Coefficients towards Robust Deep Speaker Verification}
\name{Xuechen Liu{$^1{}^,{}^2$}, Md Sahidullah{$^2$}, Tomi Kinnunen{$^1$}}
\address{
  {$^1$}School of Computing, University of Eastern Finland, Joensuu, Finland\\
  {$^2$}Universit\'{e} de Lorraine, CNRS, Inria, LORIA, F-54000, Nancy, France}

\begin{document}
%
\maketitle
\begin{abstract}
After their introduction to robust speech recognition, power normalized cepstral coefficient (PNCC) features were successfully adopted to other tasks, including speaker verification. However, as a feature extractor with long-term operations on the power spectrogram, its temporal processing and amplitude scaling steps dedicated on environmental compensation may be redundant. Further, they might suppress intrinsic speaker variations that are useful for speaker verification based on deep neural networks (DNN). Therefore, in this study, we revisit and optimize PNCCs by ablating its medium-time processor and by introducing channel energy normalization. Experimental results with a DNN-based speaker verification system indicate substantial improvement over baseline PNCCs on both in-domain and cross-domain scenarios, reflected by relatively 5.8\% and 61.2\% maximum lower equal error rate on VoxCeleb1 and VoxMovies, respectively.
\end{abstract}
\section{Introduction}
\label{sec:intro}
\emph{Automatic speaker verification} (ASV) \cite{asv2015,dnn_asv2021} aims at verifying speaker's identity from a speech waveform. To this end, an ASV system extracts acoustic features used in speaker comparison. \emph{Mel-frequency cepstral coefficients} (MFCCs) \cite{mfcc1980} and \emph{mel filterbank} outputs are two widely-used feature sets. The acoustic features are used to obtain a recording-level speaker embedding, such as an \emph{i-vector} \cite{ivector}. 

Thanks to extensive studies on deep neural network (DNN), the performance of ASV system has been substantially improved by replacing different parts of the system by DNN-based components. 
Improvements to speaker embeddings have been obtained through models such as \emph{deep speaker} \cite{deepspeaker}, \emph{recurrent end-to-end network} \cite{ge2e}, and \emph{x-vector} \cite{xvector2018}. DNN-based methods have been applied to scoring backend as well \cite{nplda2020} to improve upon statistical classifiers \cite{plda}.

Furthermore, \emph{end-to-end} front-end methods have been investigated in various applications, including ASV. In these approaches, the feature extraction and speaker information extraction are integrated together. Representative examples include SincNet \cite{sincnet, sincnet_separation2020} and its variations \cite{pase2019}. Even if promising results have been reported, end-to-end methods may have increased computational complexity, be more dependent on data, and be more challenging to interpret compared to traditional hand-crafted features.

Meanwhile, MFCCs obtained through \emph{short-term Fourier transform} (STFT) lack specificity to long-term speech characteristics and may lack robustness. Several alternatives have been developed for robust ASV and evaluated on the previous-generation statistical ASV methods. 
Feature extraction methods including some form of 
\emph{long-term processing} steps --- such as \emph{power-normalized cepstral coefficients} (PNCCs) \cite{pncc} --- have indicated their potential in improving robustness over short-term features. 
After their introduction, PNCCs have been studied in different tasks, including ASV \cite{pncc_ivec_2016, Xuechen_feature2020}. Nonetheless, while showing promise in speech recognition and keyword spotting tasks, 
the computational operations of PNCC are heavily dedicated to compensating environmental variations. While environment compensation is important in ASV as well, some of the PNCC operations may suppress intrinsic speaker variations useful in ASV tasks. Meanwhile, such variations have been studied earlier and demonstrated its efficacy in deep speaker embeddings \cite{Xuechen_feature2020}.

With the intention of developing more proper compensation techniques for ASV feature extraction, we leverage from the PNCCs as a case study. We extend its use with recent data and framework as a starting point, comparing it to MFCCs. We first try to simplify its medium-time processing module, with hypothesis on its redundancy and negative impact on ASV performance. Furthermore, we address the usefulness of channel-wise robust normalization techniques, compensating the potential loss and providing more robustness. To the best of our knowledge, this is the first work on introducing and optimizing PNCCs for speaker verification systems using deep speaker embeddings.

\begin{figure*}[t]
  \centering
  \centerline{\includegraphics[width=0.9\textwidth]{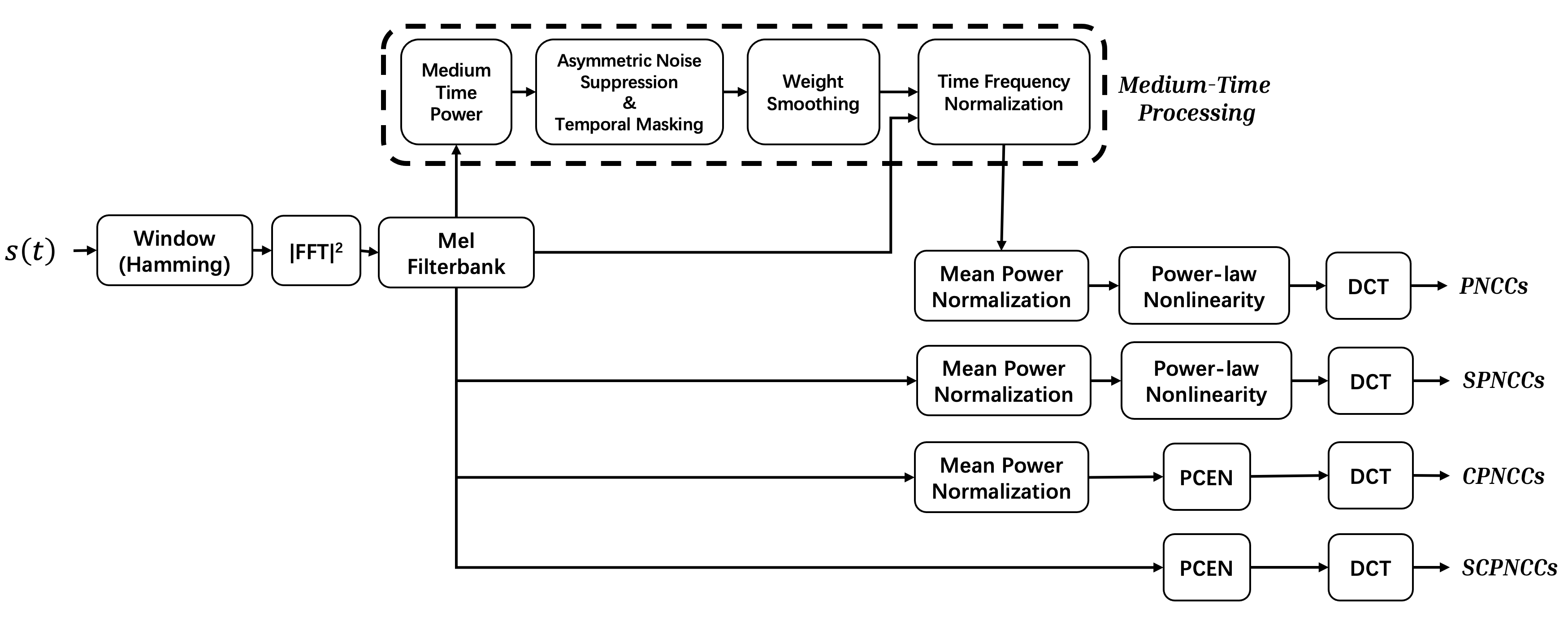}}
  \caption{Block diagram of PNCC and proposed feature extractors. PNCCs are adopted from \cite{pncc}. SPNCCs = Simple PNCCs \cite{spncc2012}; CPNCCs = Channel-normalized PNCCs; SCPNCCs = Simple Channel-normalized PNCCs. }
  \label{fig:pncc_diagram}
\end{figure*}

\begin{figure}[ht] 
  \begin{subfigure}[b]{0.5\linewidth}
    \centering
    \includegraphics[width=\linewidth]{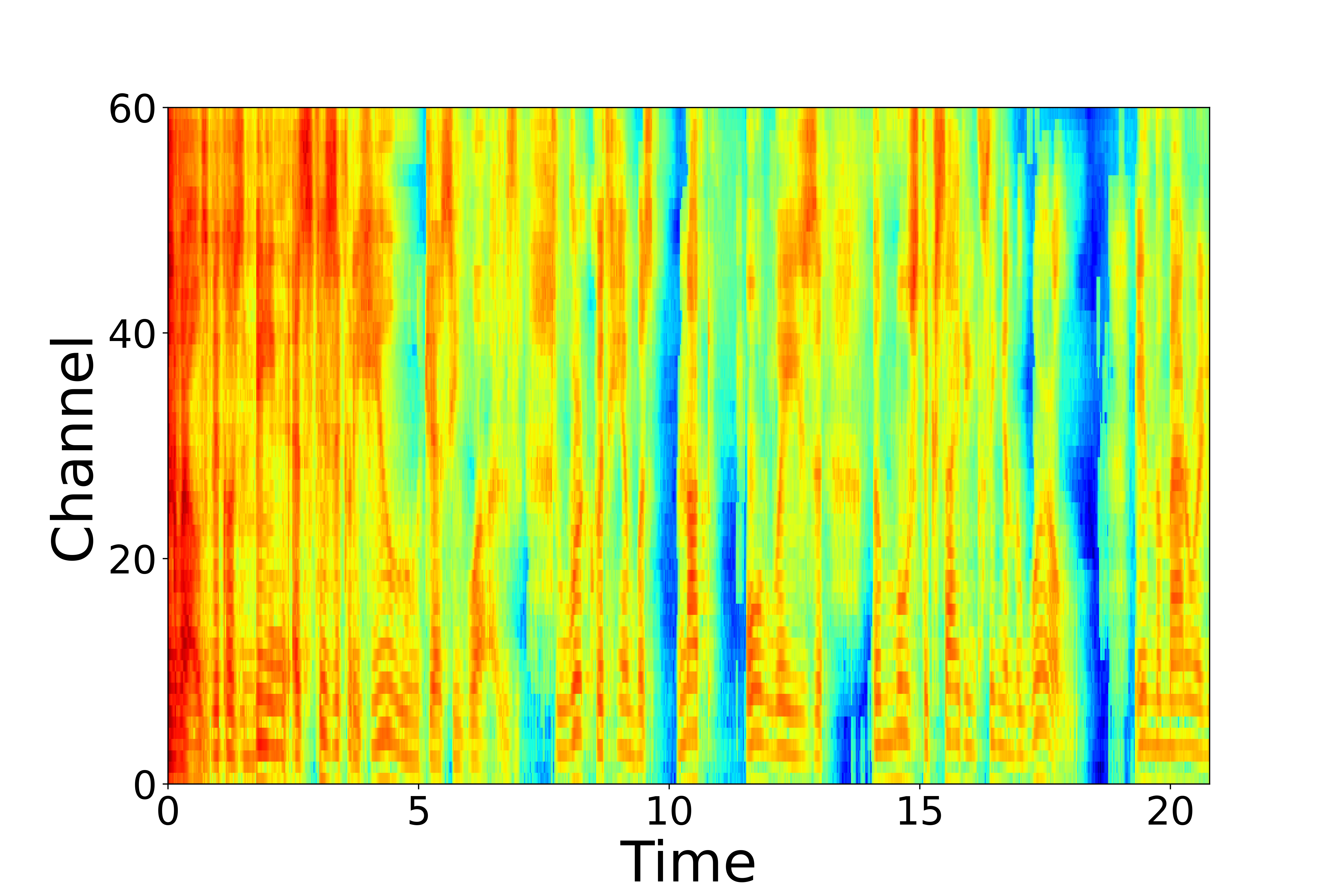} 
    \caption{PNCCs} 
    \label{fig7:pncc} 
  \end{subfigure}
  \begin{subfigure}[b]{0.5\linewidth}
    \centering
    \includegraphics[width=\linewidth]{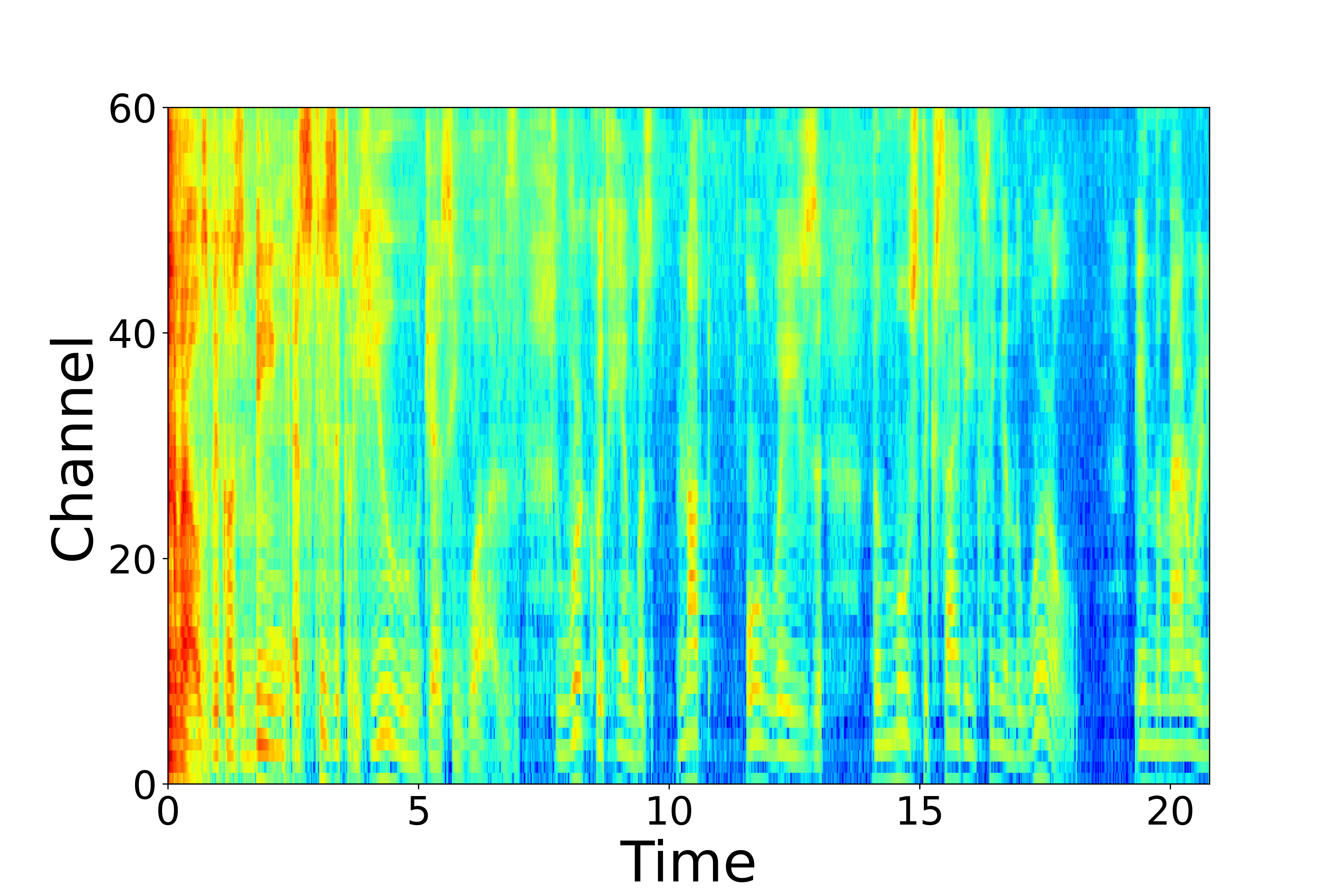} 
    \caption{SPNCCs} 
    \label{fig7:spncc} 
  \end{subfigure} 
  \begin{subfigure}[b]{0.5\linewidth}
    \centering
    \includegraphics[width=\linewidth]{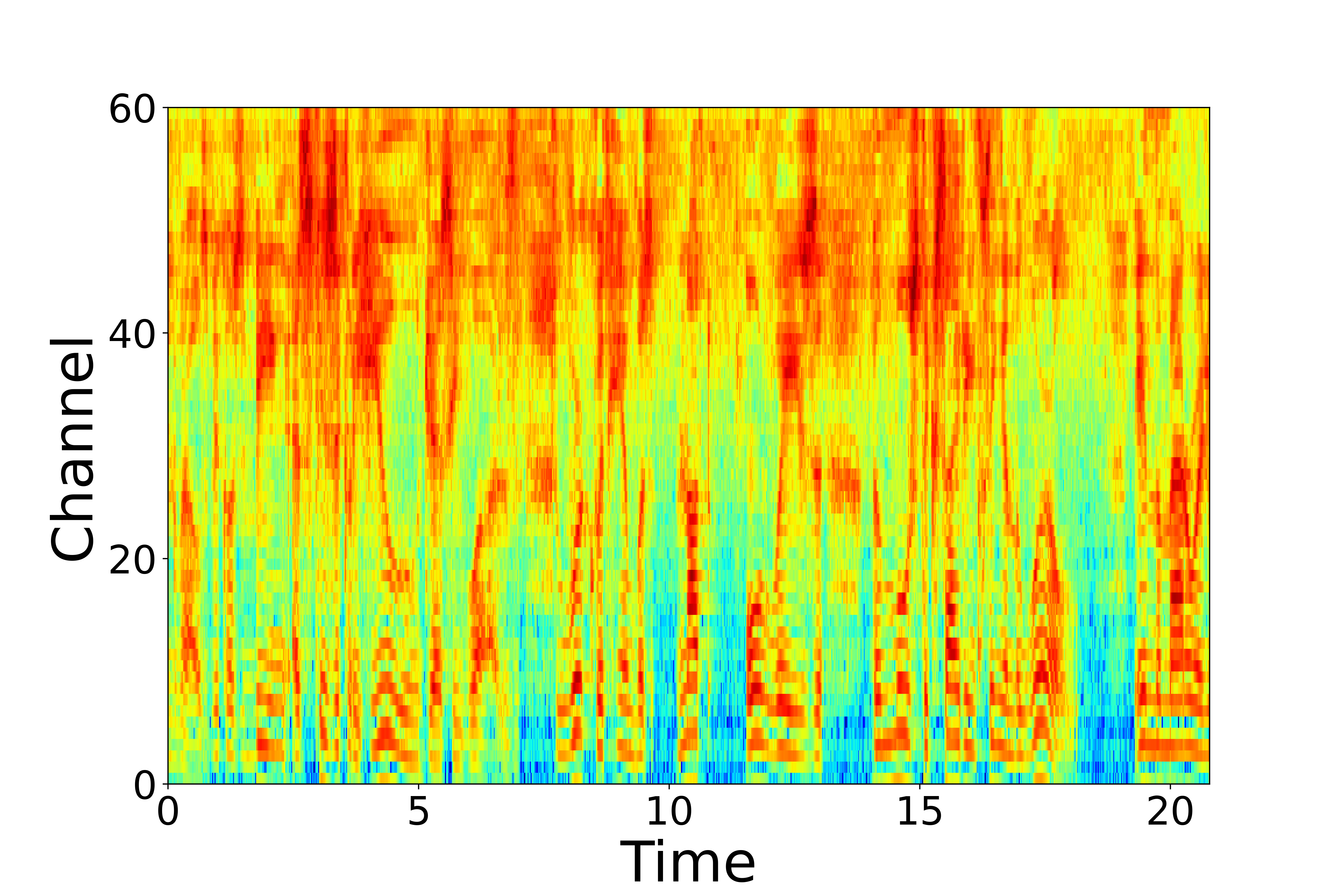} 
    \caption{CPNCCs} 
    \label{fig7:spncc_nomean_pcen}
  \end{subfigure}
  \begin{subfigure}[b]{0.5\linewidth}
    \centering
    \includegraphics[width=\linewidth]{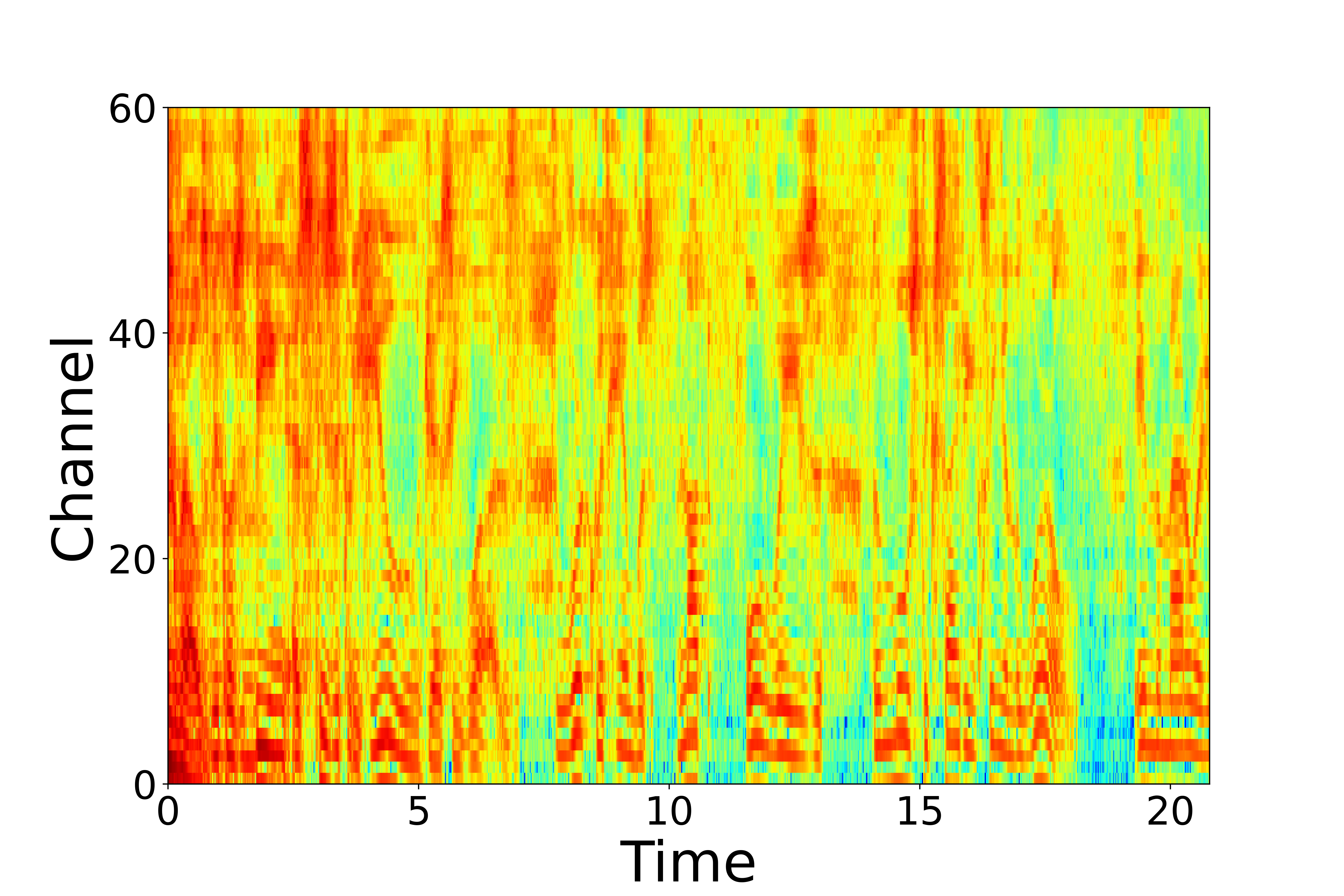} 
    \caption{SCPNCCs} 
    \label{fig7:spncc_pcen} 
  \end{subfigure}
  \caption{Effect of different feature extractors on mel spectrogram representation on a audio sample from VoxCeleb1. Number of filters is 60 for all cases. No DCT applied.}
  \label{fig:pncc_spec} 
\end{figure}

\section{Power Normalized Cepstral Coefficients}
The full pipeline of PNCCs \cite{pncc} is illustrated in Fig. \ref{fig:pncc_diagram}. Its original design is motivated by the practical need on robust features with respect to undesirable acoustic variations. Compared with established feature sets such as MFCCs \cite{mfcc1980} and RASTA-PLPs \cite{rasta}, PNCCs introduce additional signal enhancement and noise compensation operations on the spectrogram. As illustrated, those operations are represented as four steps: i) medium-time power analysis, ii) asymmetric noise suppression and temporal masking, iii) weight smoothing, and iv) time-frequency normalization. These operations are collectively referred to as \emph{medium-time processing} \cite{pncc_2012} to distinguish them from conventional short-term processing operations such as the Fourier transform. 

We consider time-frequency normalization as part of medium-time processing since it can be simply expressed as a dot-product: $\widetilde{\boldsymbol{E}}=\boldsymbol{E} \cdot \widetilde{\boldsymbol{S}}$, where $\boldsymbol{E}$ represents the short-time spectral energies after mel integration and $\boldsymbol{\widetilde{S}}$ denotes the output transfer function from weight smoothing. It is followed by mean power normalization, inspired from \emph{automatic gain control} (AGC) \cite{agc} theory. Also, the logarithmic nonlinearity in MFCC processing is replaced by a power-law nonlinearity before discrete cosine transform (DCT). In the original PNCC study \cite{pncc}, the authors also promote the usage of Gammatone filterbanks \cite{Gammatone} by noting that it may have small impact on performance compared with mel integration. Those remarks are supported by their pilot experiments. We use mel filterbank outputs for all experiments in this study.


\section{Optimization strategies on PNCCs}

In this section we describe our PNCC optimization strategies for robust deep ASV systems. First, we ablate the medium-time processing to obtain the \emph{simple power-normalized cepstral coefficients} (SPNCCs) \cite{spncc2012}. We then measure the efficacy of \emph{per-channel energy normalization} (PCEN) \cite{pcen_2017, pcen_2018}. PCEN is designed to either replace, or to jointly operate with mean power normalization. These operations act as normalizers on the channel variations and other intrinsic variations.

\subsection{Ablation of Medium-time Processing}
The medium-time processor is intended for compensating potential environmental degradation, by producing a smoothed spectral envelope. Weight smoothing, in turn, helps to suppress small perturbations. These operations can enhance speech recognition (ASR) performance \cite{pncc_2012}. However, for modern DNN-based ASV some of these operations are questionable. In particular, the smooth spectral envelope may suppress speaker-related cues, especially in voiced regions. This raises questions on the necessity of the noise suppression and smoothing operations.

Removing the medium-time module results in the SPNCCs, which has been applied earlier to robust ASR \cite{spncc2012}. Because of mean power normalization, the residual pattern with respect to time and frequency can be preserved. This can be seen in Fig. \ref{fig:pncc_spec} as well, from the perspective of time-frequency representations --- the contrast between the onsets and offsets remains. This motivates us to re-evaluate SPNCCs in robust ASV.

\subsection{Mean Power Normalization}
The module right after the medium-time processor is mean-power normalization. The mean power estimate is computed by $\mu[t] = \lambda_{\mu}\mu[t-1]+(1-\lambda_{\mu})\sum_{f=0}^{F-1}E[t,f]/L$, where $t$ and $f$ are, respectively, the time and frequency indices. $F$ denotes number of frequency bins. $E[t,f]$ represents power spectral energies, either directly from mel integration or after medium-time processing. $\lambda_{\mu}=0.999$ is known as \emph{forgetting factor}, resembling the smoothing factor in infinite impulse response (IIR) filtering. The mean power estimate is a running average of the current and preceding frame energies. The normalized power is then calculated as $\widetilde{E}[t,f]=E[t,f]/\mu[t]$.

As noted above, mean power normalization has the effect of scaling the power spectrum while retaining the local contrast. This echoes the main aim of developing such module in PNCC, which is to compensate the effect of power-law nonlinearity. However, what also can be indicated from the figures is that there are useful speech onsets that have been normalized. This may due to the power law scaling as discussed, which motivates the usage of advanced parameterized methods to properly normalize the spectrogram energies. As one of the methods whose original version have been validated effective for various speech tasks \cite{pcen_2017, pcen_dcase_2021}, PCEN is discussed in next sub-section.

\subsection{Per-channel Energy Normalization (PCEN)}
As mentioned, PCEN was developed originally as a replacement of logarithmic nonlinearity. The PCEN operation is expressed as:

\begin{align}
    \text{PCEN}[t,f] = \left(\frac{E[t,f]}{(M[t,f] + \epsilon)^{\alpha}} + \delta\right)^{r} - \delta^{r},
\label{eq:pcen}
\end{align}
where $E[t,f]$ represents the power spectral energies, $M[t,f] = (1-s)M[t-1,f] + sE[t,f]$ is the smoothed energies using an auto-regressive filter. Here we set $s$ as the reciprocal of number of frequency bins. $\alpha \in (0, 1]$ is the main parameter for compressing smoothed energies. $\delta$ and $r$ parameterize dynamic range compression, mainly aiming for properly modeling the loudness variations.

In its original study \cite{pcen_2017}, the authors treat PCEN as a replacement for logarithmic nonlinearity to address the problem of singularities and time-varying loudness. Therefore, when integrated to PNCCs, PCEN is regarded as either a replacement of power-law nonlinearity, or mean power normalization and nonlinearity. We implement both cases, resulting in \emph{channel-normalized power cepstral coefficients} (CPNCCs) and \emph{simple channel-normalized power cepstral coefficients} (SCPNCCs). Their effect is illustrated in Fig. \ref{fig7:spncc_nomean_pcen} and \ref{fig7:spncc_pcen}. Compared to PNCCs and SPNCCs, we can see that by applying PCEN, the spectrogram is enhanced, especially for higher frequency components. Meanwhile, CPNCCs preserve relatively more offsets at lower frequency bins and cast more enhancing effect relatively at higher ones than SCPNCCs. This may indicate the instability of AGC operation in PCEN, which can be compensated by placing mean power normalization as a predecessor.

\begin{table}[htbp]
  \centering
  \begin{tabular}{|c|c|c|c|c|}
    \hline
    Eval sets & Positive. & Negative. & \#target & \#non-target \\ \hline
    E-1 & D-M (\textit{same}) & D-M & 10286 & 10286 \\ \hline
    E-2 & D-I, D-M & D-I, D-M & 23289 & 23289 \\ \hline
    E-3 & D-I, D-M & D-I & 23402 & 23402 \\ \hline
    E-4 & D-I, D-M & D-M & 23433 & 23433 \\ \hline
    E-5 & D-M (\textit{diff}) & D-M & 20545 & 20545 \\ \hline
  \end{tabular}
\caption{Trial statistics of VoxMovies, taken from \cite{voxmovies}. The speech material originates from two domains: interview videos (D-I) and movie clips (D-M). Enroll and test files may come from same movie (\textit{same}) or different movies (\textit{diff}).}
\label{tab:voxmovies}
\end{table}

\section{Experiments}

In this section, we describe the experimental protocol on multiple descendants of the VoxCeleb datasets \cite{voxceleb1, voxceleb2}, with varied level of difficulty. We analyze the performance of different optimized PNCCs in both in-domain and cross-domain scenarios.

\subsection{Data}
\textbf{Training}. For all experiments we acquire speech data from the \emph{dev} of VoxCeleb2 \cite{voxceleb2} to train DNN speaker embedding extractors. It is sourced from online interview videos and consists of 5994 speakers in total. We perform data augmentation by creating copies using room impulse response (RIR) \cite{rir} and multiple sound sources from the MUSAN corpus \cite{musan}, including noise, babble and music. 

\textbf{Evaluation.} We follow the official protocols of the VoxCeleb speaker recognition challenge 2019 \cite{voxsrc2019}, generating two trials from VoxCeleb1 \cite{voxceleb1} and denoted as \emph{VoxCeleb1-E} and \emph{VoxCeleb1-H}, representing the \emph{in-domain} scenario. We also include the recent VoxMovies \cite{voxmovies} corpus in our experiments. It is sourced from movie clips with speakers selected from the VoxCeleb datasets. This creates an interesting \emph{cross-domain} speaker verification scenario. There are different trials defined based on whether audio source originates from the interviews or movie clips. There are five different trial lists, marked as E-1 to E-5. Their statistics are summarized in Table \ref{tab:voxmovies}. More details about such dataset can be found in \cite{voxmovies}, according to which the level of mismatch and verification difficulty increases from E-1 to E-5. The trial list of \emph{pooled} condition is simply the union of the five trial lists.

\subsection{System Configuration}
For all the experiments, the dimensionality of the acoustic features is set as 30, including MFCCs, without dynamic features. For all systems with the power-law nonlinearity, we set the scaling factor to $1/15$, following original PNCCs. Extended x-vector based on time-delay neural network (TDNN) \cite{Snyder_etdnn_2019} with attentive statistics pooling \cite{astats_pooling} and angular margin softmax training loss \cite{aam_softmax} forms the speaker embedding extractor. Adam \cite{adam} optimizer is applied for training, with minibatch size of 128. All speaker embeddings are extracted from the first fully-connected layer after the pooling layer. They are subsequently length-normalized and centered, prior to being transformed by a linear discriminant analyzer (LDA). Probabilistic linear discriminant analysis (PLDA) \cite{plda} is applied to produce log-likelihood scores.

\subsection{Evaluation}
We employ equal error rate (EER) and minimum detection cost function (minDCF) to evaluate ASV performance. MinDCFs were computed with target speaker prior $p_{tar} = 0.01$ and detection costs $C_\text{fa} = C_\text{miss} = 1.0$.

\begin{table}[htbp]
  \normalsize
  \centering
  \begin{tabular}{|c|cc|cc|}
    \hline
    & \multicolumn{2}{|c|}{VoxCeleb1-E} & \multicolumn{2}{|c|}{VoxCeleb1-H} \\ \hline
    Feature & EER(\%) & minDCF & EER(\%) & minDCF \\ \hline
    MFCCs & \textbf{1.88} & 0.1883 & 3.74 & 0.5808 \\ \hline
    PNCCs & 2.39 & 0.1524 & 3.91 & 0.6428 \\ \hline
    SPNCCs & 1.92 & 0.142 & 3.77 & 0.5721 \\ \hline
    CPNCCs & 2.08 & \textbf{0.0867} & \textbf{3.52} & \textbf{0.5698} \\ \hline
    SCPNCCs & 2.22 & 0.2874 & 3.84 & 0.5993 \\ \hline
  \end{tabular}
\caption{Results on VoxCeleb1 evaluation sets.}
\label{tab:vox1_results}
\end{table}

\begin{figure}[h]
  \centering
  \centerline{\includegraphics[width=0.55\textwidth]{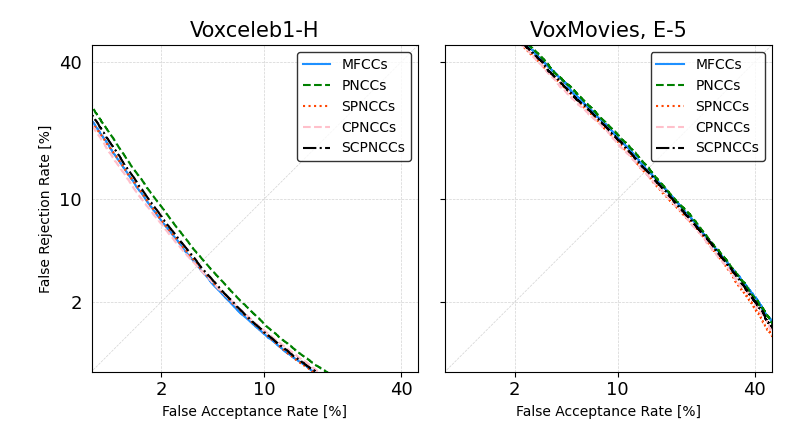}}
  \caption{DET visualization of systems with different feature extractors on \emph{VoxCeleb1-H} (left) and E-5 from VoxMovies (right). Best viewed in color.}
  \label{fig:pncc}
\end{figure}

\section{Speaker Verification Results}

\subsection{Voxceleb1}
The results on VoxCeleb1 reported in table \ref{tab:vox1_results} indicate that the performance of the feature extractors differs between the two conditions.
Let us first look at \emph{VoxCeleb1-E}, the least challenging condition across all covered in this study. While CPNCCs achieve the lowest minDCF among all systems, in terms of EERs none of the proposed features outperform conventional MFCCs. This confirms the efficacy of MFCCs under in-domain or relatively matched conditions. Among PNCC variants, the lowest EER is obtained from SPNCCs.

We then move on to the more challenging \emph{VoxCeleb1-H}. While standard PNCCs still remain behind MFCCs, removing the medium-time processing component provides a 3.6\% relative improvement in EER. Applying PCEN furthers such improvement and results in best performance among all features in both metrics, outperforming MFCCs by relatively 5.8\% in EER and 18.9\% in minDCF. These results indicate the robustness provided by PCEN in the more challenging \emph{VoxCeleb1-H} condition. Meanwhile, removing mean power normalization, again degrades performance. Mean power normalization thus is speculated to have an improvement role in normalizing in-domain acoustic variations.

\begin{table}[htbp]
  \centering
  \begin{tabular}{|c|cccccc|}
    \hline
    Feature & \emph{Pooled} & E-1 & E-2 & E-3 & E-4 & E-5 \\ \hline
    MFCCs & 13.15 & 11.04 & 12.04 & 9.22 & 18.93 & 14.43  \\ \hline
    PNCCs & 13.72 & 10.70 & 12.84 & 6.79 & 20.17 & 14.76 \\ \hline
    SPNCCs & 13.09 & \textbf{10.14} & 12.04 & 4.91 & 19.53 & 13.08 \\ \hline
    CPNCCs & \textbf{13.0} & 10.39 & \textbf{11.72} & \textbf{3.57} & \textbf{18.3} & \textbf{13.03} \\ \hline
    SCPNCCs & 13.30 & 10.30 & 12.20 & 5.65 & 18.42 & 14.36 \\ \hline
  \end{tabular}
\caption{EER(\%) results on VoxMovies.}
\label{tab:voxmovies_results}
\end{table}

\subsection{VoxMovies}

Performance of different feature extractors on VoxMovies trial sets in terms of EER is presented in table \ref{tab:voxmovies_results}. The standard PNCCs outperforms MFCCs slightly on first three conditions but not on E-4 and E-5. Removing medium-time processing results in improvement across all conditions, including \emph{pooled}. VoxMovies contains more intrinsic speaker variations and domain mismatch compared to the VoxCeleb1 conditions. Therefore, based on results from MFCCs, PNCCs and SPNCCs, medium-time processing is suspected to have a negative impact.

Moving on to the two variants with PCEN, best performance on all conditions and \emph{pooled}, excluding E-1, is achieved by CPNCCs, with maximum relative improvement of 61.2\% observed in the E-3 condition. Since E-1 contains least domain mismatch according to Table \ref{tab:voxmovies}, this result is expected. For the most challenging E-5 it reaches 9.5\% relative improvement in EER over MFCCs. Therefore, integrating mean power normalization and PCEN thus may have positive impact on normalizing the intrinsic variations, either as an addition or replacement from the perspective of medium-time processing. Ablating mean power normalization gain (SCPNCCs) leads to performance drop on all conditions except E-1, but this time SCPNCCs outperforms MFCCs on four out of the five conditions. This indicates the efficacy of channel normalization on reducing domain mismatch, without requiring extensive parameter tuning. Future work thus can focus on domain adaptation on related parameters via fine-tuning or making data-adaptive version of PCEN, for either improving CPNCCs or refining SCPNCCs, as described in \cite{pcen_2017}.

\subsection{Analysis with DET plot}
We analyze the performance of different feature extractors on the two challenging conditions: \emph{VoxCeleb1-H} and E-5 from VoxMovies. For \emph{VoxCeleb1-H}, we can observe that while the gain from CPNCCs are reflected, for scenario where systems are less strict on false alarms (right bottom), MFCCs may be a better choice over the proposed PNCC variants. This is different from what we see at the other plot, where the gain from CPNCCs and SPNCCs over other systems are consistent, echoing what has been reported via EER.

\section{Conclusions}
We have re-evaluated and optimized power normalized cepstral coefficients (PNCCs) for deep neural network based speaker verification. We first ablated the medium-time processing module, which results in simple PNCCs (SPNCCs) and indicates promising performance. We did further simplification by applying PCEN. 
We performed experiments in both in-domain (VoxCeleb1) and cross-domain (VoxMovies) scenarios. Among all the observed improvements, we highlight particularly the successful combination of mean power normalization and PCEN as a replacement of power-law nonlinearity and medium-time processing (CPNCCs). In the cross-domain scenario, the relative maximum improvement over the baseline was 61.2\%.

In future work, we may focus on creating data-adaptive or pre-trained variants with optionally sequential training criteria, similar to what has been done in \cite{pcen_2017}. Exploring individual PCEN components (AGC and DRC) as alternative choices from nonlinearities such as logarithmic and power-law is also left as future work.

\section{Acknowledgements}
This work was partially supported by Academy of Finland (project 309629) and Inria Nancy Grand Est.

\bibliographystyle{IEEEbib}
\bibliography{strings,refs}

\end{document}